\documentclass[aps,prl,floats,twocolumn,showpacs,superscriptaddress]{revtex4}

\usepackage{epsfig}
\usepackage{latexsym,amsmath}

\begin{document}
\title{Percolation and Epidemic Thresholds in Clustered Networks}

\author{M. \'Angeles Serrano}

\affiliation{School of Informatics, Indiana University,\\ Eigenmann
Hall, 1900 East Tenth Street, Bloomington, IN 47406, USA}

\author{Mari{\'a}n Bogu{\~n}{\'a}}

\affiliation{Departament de F{\'\i}sica Fonamental, Universitat de
  Barcelona,\\ Mart\'{\i} i Franqu\`es 1, 08028 Barcelona, Spain}

\date{\today}

\begin{abstract}
We develop a theoretical approach to percolation in random clustered
networks. We find that, although clustering in scale-free networks
can strongly affect some percolation properties, such as the size
and the resilience of the giant connected component, it cannot
restore a finite percolation threshold. In turn, this implies the
absence of an epidemic threshold in this class of networks
extending, thus, this result to a wide variety of real scale-free
networks which shows a high level of transitivity. Our findings are in good agreement with numerical simulations.

\end{abstract}


\maketitle

Perhaps one of the main reasons for the growing interest in complex
networks is that, indeed, many systems in the real world, either
naturally evolved or artificially designed, are organized in a
networked fashion~\cite{Albert:2002,Mendesbook}. This makes any
theoretical approach potentially applicable to many different fields
in the short term. As a germane example, percolation on
networks has been one of these theoretical advances which has helped
to understand, for instance, the high resilience of scale-free (SF)
networks in front of the removal of a fraction of their
constituents, with important implications for communication systems
like the Internet and other Peer-To-Peer networks~\cite{RomusVespasbook}.

In addition to its high theoretical interest, percolation theory
serves as a conceptual framework to treat more factual problems on
networks, such as the dynamics of epidemic spreading~\cite{Anderson:1991}. Indeed, the
susceptible-infected-removed (SIR) model of epidemic spreading can
be mapped into a bond percolation problem~\cite{Grassberger:1983,Sander:2002,Sander:2003,Newman:2002b}.
This is one of the simplest models in the
literature~\cite{Kermack:1927,RomusVespasbook2002}, with three different states for the
elements of the population: susceptible, infected, and removed. In
its bare formulation, it is characterized by the time that an
individual remains infected and the time that an infected individual
takes to infect a susceptible neighbor, both random variables
following a Poisson process but with different constant rates. Since
the infection uses the network as a template to spread, the process
of propagation can be understood as a percolation problem over the
original network where each edge is removed with probability
$q_{inf}=1-p_{inf}$, being $p_{inf}$ the likelihood that an infected
individual infects a susceptible neighbor before becoming removed.
This mapping stands as an example of the importance of percolation
theory beyond theoretical concerns.

Percolation properties of random directed and undirected networks
with given degree distributions and two-point correlations have been
extensively studied
\cite{Havlin:2000,Havlin:2002,Newman:2002a,Vazquez:2003,Boguna:2005}. One of the
most striking results, due to its important implications, is the
absence of a percolation threshold in uncorrelated random SF
networks~\cite{Romualdo:2001,Havlin:2000}. In other words, in this
type of networks, one has to remove virtually the totality of their
constituents before the network fragments into disconnected
components. Translated into the epidemic context, this means that
an epidemic threshold below which the epidemics
cannot propagate does not exist. This result is particularly important due to the
fact that a large number of real networks have a SF degree
distribution. This result has also been generalized to the case of
random SF networks with two-point correlations, both for the
SIR model and for the susceptible-infected-susceptible (SIS) model
of epidemic spreading~\cite{Boguna:2003,Vazquez:2003}.

Nevertheless, almost all the analytical results obtained up to date
implicitly refer to networks without clustering and little is known about its effects on the percolation properties of such networks, with the exception of Ref.~\cite{Newman:2003}, where an analytical solution for the percolation properties of the one-mode projection of random bipartite graphs was developed. See also \cite{Volz:2004}. This is due to the fact that those analysis are based on the idea of branching process. 
This approach works well when the network is locally tree-like
and, thus, the clustering coefficient is very small. Real networks,
however, are shown to have a significant level of clustering that may change the percolation properties
significantly. In this paper, we present analytical and simulation
results for percolation in clustered networks. The analytical approximation becomes exact in the limit of weak clustering and simulations are also provided in the case of strong clustering. We find that clustering makes networks more fragmented as compared to the unclustered counterparts but with giant components which have tighter interconnected cores of high-degree vertices. We also find that {\it clustering cannot restore the percolation and epidemic thresholds in SF networks}.

To begin with, we follow Ref.~\cite{Serrano:2005b} and define the
multiplicity of an edge, $m_{ij}$, as the number of triangles in
which the edge connecting vertices $i$ and $j$ participates. This
quantity is the analog to the number of triangles attached to a node
$i$, $T_{i}$, which is used to define the local clustering
coefficient. In the coarse-grained level of degree classes, one can define the
multiplicity matrix $m_{kk'}$ as the average multiplicity of the
edges connecting the classes $k$ and $k'$. Then, the
degree-dependent clustering coefficient $\bar{c}(k)$ --a property of
vertices-- and the multiplicity matrix $m_{kk'}$ --a property of
edges-- are related through the following identity valid for any
network
\begin{equation}
\sum_{k'} m_{kk'}P(k,k') = k(k-1) \bar{c}(k) \frac{P(k)}{\langle k
\rangle}, \label{db}
\end{equation}
where $P(k)$ is the degree distribution and $P(k,k')$ is the probability that one edge connects two vertices of degrees $k$ and $k'$. The multiplicity matrix $m_{kk'}$, which varies in the range
$[0,m^c_{kk'}]$ with $m^c_{kk'}=min(k,k')-1$, gives a more detailed
description on how triangles are shared among vertices of different
degrees and, as we shall see, it contains the relevant information
to analyze the percolation properties of clustered networks.

An alternative way to quantify clustering is by using the edge
clustering coefficient as defined in~\cite{Radicchi:2004}
\begin{equation}
\bar{c}(k,k')=\frac{m_{kk'}}{min(k,k')-1}.
\end{equation}
As in the case of the local clustering coefficient, $\bar{c}(k,k')$
also has a probabilistic interpretation. It quantifies the
likelihood that a pair of connected vertices have a common neighbor.
If the network is random, we can assume that the probability that an
edge connecting two vertices of degrees $k$ and $k'$ has
multiplicity $m$ is
\begin{equation}
\phi(m|kk')=\left(
\begin{array}{c}
m^c_{kk'}\\
m
\end{array}
\right) [\bar{c}(k,k')]^m[1-\bar{c}(k,k')]^{m^c_{kk'}-m}.
\end{equation}
This probability, along with the multiplicity matrix, are crucial to
compute correctly the percolation properties of clustered random
networks due to the fact that, although we start from a given vertex
and we follow all its edges as in the non-clustered case, once we
are placed in one of the neighbors, we only follow those edges not
pointing to the neighborhood of the source vertex so that we avoid
edges responsible for clustering. It is worth noticing that, even in
this scheme, we are neglecting the fact that higher order loops may
be present.

Let us start the analytical computations by defining the probability
that a given vertex has $s$ reachable vertices (including itself),
$G(s)$. For very heterogeneous networks it is more convenient to
define this probability conditioned to the degree of the source
vertex, $G(s|k)$, and then $G(s)=\sum_k P(k) G(s|k)$. Finally, we need to introduce an extra
function, $g(s|k)$, which measures the probability that a vertex can
reach $s$ other vertices given that it is connected to a vertex $v$,
of degree $k$, and that it cannot visit neither $v$ nor its
neighborhood (this idea was used in \cite{Newman:2003b} to compute the number of second neighbors of a given vertex). This last condition guaranties that we do not
overcount contributions due to triangles. The functions $G(s|k)$ and
$g(s|k)$ are related through
\begin{equation}
G(s|k)=\sum_{s_1,\cdots,s_k}g(s_1|k) \cdots g(s_{k}|k)
\delta_{s,1+s_1+\cdots +s_{k}}.
\end{equation}
We can find a recursion relation for $g(s|k)$ taking into account
that now the branching process has the constraint that at each
generation point we can only use the free edges to continue the
exploration. In this case
\begin{eqnarray}
g(s|k)=\sum_{k'} \sum_m P(k'|k) \phi(m|k,k')\nonumber \\
\sum_{s_1,\cdots}g(s_1|k') \cdots g(s_{k'_{br}}|k')
\delta_{s,1+s_1+\cdots +s_{k'_{br}}}, \label{gsk}
\end{eqnarray}
where $k'_{br}=k'-m-1$. To simplify this equation we make use of the
so-called generating function formalism and transform $g(s|k)$ to
the discrete Laplace space, $\hat{g}(z|k)\equiv \sum_s z^s g(s|k)$,
where Eq.~(\ref{gsk}) becomes a closed equation for the function
$\hat{g}(z|k)$,
\begin{equation}
\hat{g}(z|k)=z \sum_{k'} \sum_m P(k'|k) \phi(m|k,k')
\left[\hat{g}(z|k')\right]^{k'_{br}}. \label{gzk}
\end{equation}
The percolation transition takes place when Eq.~(\ref{gzk}),
evaluated at $z=1$, admits as a stable solution $\hat{g}(z=1|k)=\xi
(k)\le 1$, that is, there is a finite probability ($1-\xi(k)$) that
the branching process extends up to infinity. To analyze the
stability of Eq.~(\ref{gzk}) near the fixed point $\hat{g}(z=1|k)=1$
we study a perturbative solution $\hat{g}(z=1|k)\approx
1+\chi(k)\epsilon$ in the limit $\epsilon \rightarrow 0$. From
Eq.~(\ref{gzk}),
\begin{equation}
\chi(k)=\sum_{k'} (k'-1-m_{kk'}) P(k'|k)  \chi(k'), \label{chi}
\end{equation}
using that $m_{kk'}=\sum_m m\phi(m|k,k')$.
The transition between the percolated and the fragmented phases is
given by the properties of the matrix $(k'-1-m_{kk'}) P(k'|k)$, and,
in particular, by its maximum eigenvalue $\Lambda_m$. When
$\Lambda_m>1$ the network is in the percolated phase in which a
macroscopic fraction of the system becomes globally connected. In
the opposite situation, the network is a set of small disconnected
clusters.
\begin{figure}[t]
\epsfig{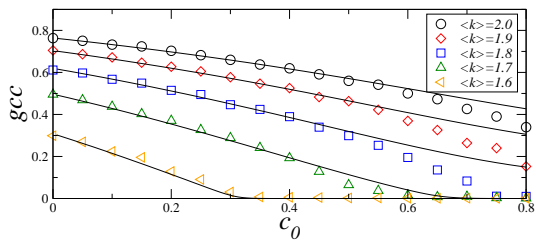}
 \caption{Relative size of the giant component as a function of
 $c_0$, $\bar{c}(k)=c_0/(k-1)$, for different average degrees and an exponential degree distribution for networks generated with the
algorithm of Ref.~\cite{Serrano:2005b} (network size is $N=10^5$). Solid
lines correspond to the numerical solution of Eq.~(\ref{gzk}). In the case $c_0=0$, we recover the results of the configuration model.}
\label{giant1}
\end{figure}

The simplest case of clustered network corresponds to $m_{kk'}=m_0$,
with $m_0 \in [0,1]$. In this
situation, from Eq.(\ref{db}) one obtains $\bar{c}(k)=c_0
(k-1)^{-1}$, where $c_0$ is a function of $m_0$ to be determined.
Hence, small degree nodes are highly clustered whereas high degree
ones are less clustered. This specific form of $\bar{c}(k)$ is
particularly important since it represents the maximum level of
clustering one can impose in a network without introducing at the
same time degree-degree correlations. This will allow us to analyze
the effect of triangles without any interference from two-point
correlations. Hereafter, we will refer to levels of clustering below
this threshold as {\it weak transitivity}. In fact, two-point
correlations can be totally avoided except for vertices of degree
$k=1$ --that do not participate in triangles-- and must necessarily
follow a different connection pattern. The clustering factor $c_0(m_0)$ takes in this case the form
\begin{equation}
c_0(m_0)=m_0\frac{1-2\frac{P(1)}{\langle k
\rangle}+P(1,1)}{(1-\frac{P(1)}{\langle k \rangle})}.
\end{equation}
The probability $P(1,1)\equiv x$ is the smallest solution of the following quadratic equation (the derivation will be given in a forthcoming publication)
\begin{equation}
x^2-\left(\frac{\langle \phi \rangle'}{1-\langle \phi \rangle'}+\frac{2P(1)}{\langle k \rangle}\right)x+\frac{P^2(1)}{\langle k \rangle^2 (1-\langle \phi \rangle')}=0
\label{eq:P(1,1)b}
\end{equation}
where $\langle \phi \rangle'$ is the average of $\phi(0|kk')$ over the set of vertices of degrees larger than 1. Then, the maximum eigenvalue of the matrix
$(k'-1-m_{kk'}) P(k'|k)$ can be analytically computed and so the
percolation condition
\begin{equation}
\frac{\langle k(k-1)\rangle}{\langle k \rangle} >(1+c_0(m_0))
\frac{m_0}{c_0(m_0)} (1-\frac{P(1)}{\langle k \rangle}). \label{cp}
\end{equation}
For very low clustering, we recover the well-known result for
percolation in random networks. The immediate conclusion seems to be
that clustering changes the position of the critical point. However,
in the case of SF networks, the left hand side of
Eq.(\ref{cp}) diverges in the thermodynamic limit and, therefore, in
SF networks {\it weak transitivity} is not able to restore a finite percolation
threshold, and hence, a finite epidemic threshold.

To check the accuracy of the present formalism, we generated
clustered random networks using the algorithm introduced in
Ref.~\cite{Serrano:2005b}. We simulated networks of $10^5$ nodes
with an exponential degree distribution and a clustering coefficient
$\bar{c}(k)=c_0 (k-1)^{-1}$. In Fig.~\ref{giant1}, we compare the
relative size of the giant connected component, $gcc$, as a function of
$c_0$ with the numerical solution of the Eq.(\ref{gzk}). As it can be
seen, the effect of clustering is to reduce the size of the giant
connected component (in agreement with \cite{Newman:2003,Volz:2004}). The effect is so strong that, in networks with
a moderate average degree, it can fragment completely the network
when $c_0$ exceeds a critical value. In other cases, the reduction
of the size can be more than $50\%$. For values of $c_0
\in [0,0.5]$, the agreement between our formalism and the numerical
simulations is excellent. Beyond this point, our approximation
slightly overestimates the $gcc$'s size. This is mainly due to the
fact that in this regime, links of multiplicity larger than 1 appear
which, in turn, induces the presence of some loops of order four.
\begin{figure}[t]
\epsfig{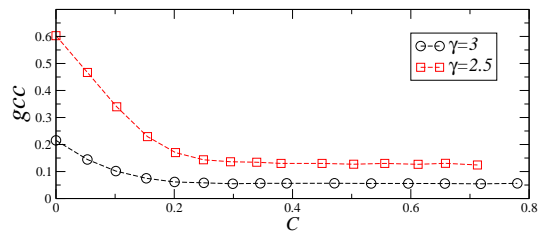}
 \caption{Relative size of the giant component
 as function of the global clustering $C$ for scale-free
 networks with $\gamma=3$ and $\gamma=2.5$ and {\it strong transitivity}.
 The giant components converge to a constant value independent
 of the level of clustering. Each point corresponds to a network of size $N=10^5$.
} \label{giant2}
\end{figure}

We now turn our attention to the case of {\it strong transitivity},
which corresponds to functions $\bar{c}(k)$ decaying slower than
$k^{-1}$. In this case, clustering and two-point degree correlations
are intimately coupled~\cite{Serrano:2005b}. An heuristic argument
is as follows: if a vertex with a high degree has also a high
clustering coefficient, many of its neighbors will be connected
among them, which induces an assortative behavior. In other words,
to generate random networks with {\it strong transitivity} we need
to introduce some mechanism generating assortativity. However, it is not
possible to obtain a perfect assortative pattern in SF
networks for arbitrary large degrees (see Ref.~\cite{Boguna:2004}
for a detailed discussion) and, as a consequence, the maximum level
of clustering is limited. The algorithm of Ref.~\cite{Serrano:2005b}
has a free parameter which allows to control the assortativity of
the resulting network so that SF networks with high
clustering can be generated. We quantify the level of clustering as
$C=(1-P(1))^{-1}\sum_{k}P(k) \bar{c}(k)$, so that $C$ is defined in
the interval $[0,1]$. In Fig.~\ref{giant2}, we show the relative
size of the giant component as a function of $C$. As in the case of
{\it weak transitivity}, clustering reduces the size of the giant
component. However, after a certain value, the size of the giant
component stabilizes to a constant value which is independent of
$C$.
\begin{figure}[t]
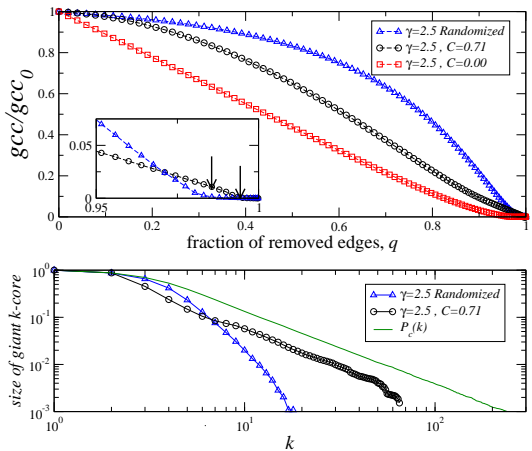

\centering $\begin{array} {c}
 \epsfig{file=fig3.eps,width=7cm} \\
  \epsfig{file=fig4.eps,width=6.92cm}
\end{array}$
 \caption{Top: Relative size of the giant component in relation
 to the original size ($N_{gcc}=601353$ for the unclustered network, and
$N_{gcc}=125353$ for the clustered one) as a function of the fraction of removed
 edges, $q$. SF nets with $\gamma=2.5$ and $N=10^6$ are simulated for: {\it i)} a clustered network,
 $C=0.71$ (circles), {\it ii)} an unclustered one (squares), and {\it iii)}
 the randomized $gcc$ of the clustered net. The inset shows a zoom of the area close to $q=1$. Bottom:
 Relative sizes of the giant $k$-cores for {\it ii)} and {\it iii)}
 and the cumulative degree distribution.
} \label{giant3}
\end{figure}
Therefore, SF networks with high levels of clustering have giant
components which are smaller than their counterparts in networks
without clustering. But, which are the resilience properties of
those giant components in front of random removal of edges? To
answer this question, we have generated two SF networks with
$\gamma=2.5$, one with the maximum level of clustering (C=0.71) and
the other without clustering, and applied a random removal of edges
on the corresponding giant components. The results are shown
in Fig.~\ref{giant3} (top graph). The giant component of the
clustered network turns out to be more resilient than the giant
component of the unclustered one. Since SF networks without
clustering does not have a percolation threshold, we conclude that
{\it clustering, even high, cannot restore the percolation and
epidemic thresholds in random SF networks}.

However, the degree distributions of the giant connected components
can be different, a fact that could explain the observed differences
in the resilience properties. To check this point, we have
randomized the $gcc$ of the clustered network while keeping
fixed its degree distribution (see the curve labeled {\it
Randomized} in the top of Fig.~\ref{giant3}). This network is more
resilient than the clustered one for all levels of damage except for
very high values, for which the $gcc$ of the randomized network goes
to zero faster due to finite size effects. This is illustrated in
the inset of Fig.~\ref{giant3}. The first arrow indicates the
threshold computed with the formula $q_{c}=1-\langle k \rangle/\langle
k(k-1)\rangle=0.986 $, whereas the second arrow indicates the
threshold due to finite size effects for the clustered net, which is
placed closer to $1$. Therefore, clustered networks
are less sensitive to finite size effects than random equivalent
ones. This can be understood analyzing the $k$-core decomposition of
the networks (see \cite{Dorogovtsev:2006} and references therein). The $k$-core is the maximal subgraph such that all its nodes have $k$ or more connections within the subgraph. In the
bottom plot of Fig.~\ref{giant3}, we show the relative size of the
giant $k$-core for both networks. For small $k$, the
randomized network has $k$-cores which are bigger than the ones of
the clustered net, which explains why it is more resilient. However,
for very large degrees, the clustered network has bigger $k$-cores,
that is, it exists a small but finite core of vertices with very
large degrees highly interconnected among them, which makes the
network less prone to finite size effects. We also show the
cumulative degree distribution $P_{c}(k)$, since  it bounds the
sizes of the $k$-cores, which, for the clustered net, decays as a function of $k$ with the
same exponent.

Summarizing,  we have introduced a theoretical framework to analyze
percolation properties of clustered networks. We have shown that,
although clustering strongly affects the percolation properties and
the sizes of the giant components {\it it cannot restore the
percolation and epidemic thresholds in random SF networks},
extending, thus, this important result to a wider class of networks,
closer to the real ones. It is also worth to mention that these results can also be applied to other epidemiological models like the SIS model.

\begin{acknowledgments}
We thank A. Vespignani and  R. Pastor-Satorras for valuable comments.
This work has been partially supported by DGES,
Grant No. FIS2004-05923-CO2-02 and Generalitat de Catalunya Grant No. SGR00889. M. B. thanks the School of Informatics at Indiana University, where part of this work was developed. 
\end{acknowledgments}


\end{document}